\documentclass{article}
\usepackage{amssymb,bbm,amsmath,stmaryrd,subfig,color,mathrsfs,graphicx}
\usepackage{pxfonts}
\usepackage[paperwidth=16cm,paperheight=24cm,textwidth=10cm,textheight=10cm,top=2cm,left=2cm,right=2cm,bottom=2cm]{geometry}
\usepackage{hyperref}
\definecolor{refcolor}{rgb}{0.3,0.3,0.7}
\hypersetup{
    pdftitle={The contact of elastic regular wavy surfaces revisited},
    pdfauthor={Vladislav A. Yastrebov},
    pdfsubject={Mechanics of rough contact},
    pdfcreator={LaTeX, Kile},
    pdfkeywords={wavy surface, elastic contact, contact area, contact perimeter, compactness of contact area, probability density of contact pressure},
    pdfnewwindow=true,
    colorlinks=true,
    linkcolor=refcolor,          
    citecolor=refcolor,          
    filecolor=refcolor,      
    urlcolor=refcolor           
}
\usepackage{graphicx}
\usepackage{xspace}
\usepackage{amssymb,amsmath}
\usepackage{marvosym}
\usepackage{color}

\newcommand{\be}{\begin{equation}}
\newcommand{\ee}{\end{equation}}
\newcommand{\bes}{\begin{equation*}}
\newcommand{\ees}{\end{equation*}}

\definecolor{greenvy}{RGB}{0,128,0}
\definecolor{bluevy}{RGB}{0,191,191}
\definecolor{violetvy}{RGB}{191,0,191}
\definecolor{orangevy}{RGB}{255,150,0}

\newcommand{\ddp}[2]{\frac{\partial{#1}}{\partial{#2}}}
\newcommand{\ddps}[2]{\frac{\partial{^2\!#1}}{\partial{#2^2}}} 

 \title{The contact of elastic regular wavy surfaces revisited}
 
 \author{Vladislav A. Yastrebov$^{a}$\footnote{Corresponding author \href{mailto:vladislav.yastrebov@mines-paristech.fr}{vladislav.yastrebov@mines-paristech.fr}}\\ Guillaume Anciaux$^b$\qquad Jean-Fran\c cois Molinari$^b$}

 \date{\footnotesize$^a${\it MINES ParisTech, PSL Research University, Centre des Mat\'eriaux, CNRS UMR 7633, BP 87, F 91003 Evry, France}\\
                    $^b${\it Civil Engineering (LSMS, IIC-ENAC, IMX-STI), Ecole Polytechnique F{\'e}d{\'e}rale de Lausanne (EPFL), Station 18, 1015, Lausanne, Switzerland}}

\usepackage{graphicx}
\usepackage{xspace}
\usepackage{amssymb,amsmath}
\usepackage{marvosym}
\begin{document}

\maketitle

\begin{abstract}
We revisit the classic problem of an elastic solid with a two-dimensional wavy surface squeezed against an elastic flat half-space from infinitesimal to full contact. 
Through extensive numerical calculations and analytic derivations, we discover previously overlooked transition regimes.
These are seen in particular in the evolution with applied load of the contact area and perimeter, the mean pressure and the probability density of contact pressure.
These transitions are correlated with the contact area shape, which is affected by long range elastic interactions. 
Our analysis has implications for general random rough surfaces, as similar local transitions occur continuously at detached areas or coalescing contact zones.
We show that the probability density of null contact pressures is non-zero at full contact.
This might suggest revisiting the conditions necessary for applying Persson's model at partial contacts and
guide the comparisons with numerical simulations.
We also address the evaluation of the contact perimeter for discrete geometries and the applicability of Westergaard's solution for three-dimensional geometries.
\end{abstract}


\tableofcontents

\section{\label{intro}Introduction}

Contact interaction between solids is affected by inevitable roughness of surfaces.
This roughness is linked with stress fluctuations, friction, adhesion and wear of contacting solids, 
as well as leakage and energy transfer through contact interfaces.
A two-dimensional sine-wave surface is one of the simplest models of periodic and continuous surface roughness.
This model has been considered in many applications, including optical scatter~\cite{stover2012b}, mechanical contact~\cite{johnson1985ijms}, heat and electric transfer~\cite{krithivasan2007tl},
fluid flow slip conditions~\cite{einzel1990prl} and others.
Although apparently simple, finding the analytical solution for the mechanical contact between an elastic solid possessing 
a double sine wave surface with an elastic plane seems improbable~\cite{johnson1985ijms}.
Since the last three decades, numerical methods allow to overpass the limitations and complexity of analytical methods and obtain
solutions for arbitrary geometries and boundary conditions.
About thirty years ago, Johnson, Greenwood and Higginson~\cite{johnson1985ijms} studied in detail the contact between elastic regular wavy surfaces.
Although no closed-form analytic solution was obtained, the question of squeezing a wavy surface was considered fully addressed and the results of~\cite{johnson1985ijms} served as a reference for verification of following numerical
studies, see e.g.~\cite{krithivasan2007tl}.

\begin{figure}
 \includegraphics[width=1\columnwidth]{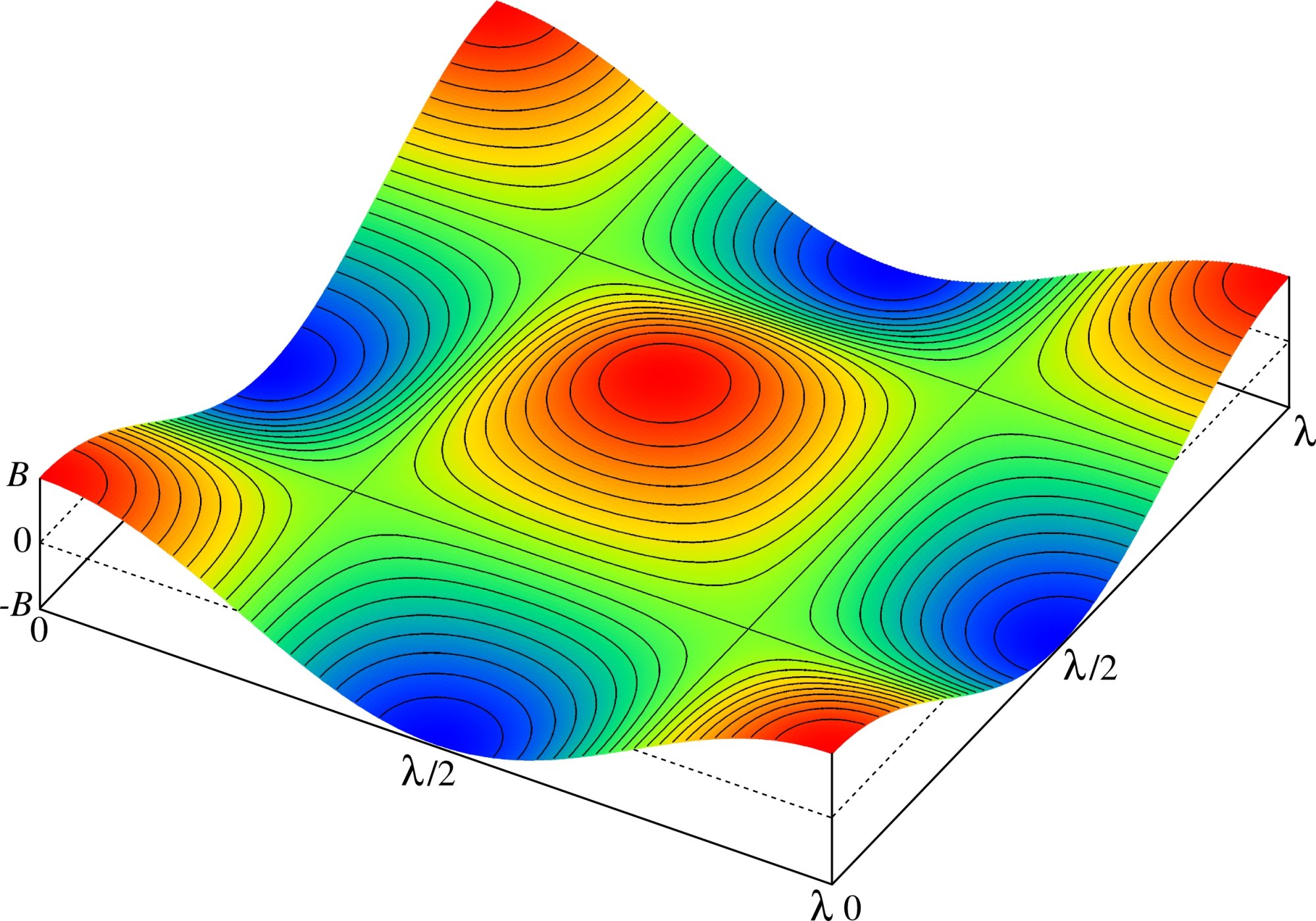}
 \caption{\label{fig:1}Simulation domain with a double sine wave surface, Eq.~\eqref{eq:surface}}
\end{figure}

In this paper, we take a fresh look at this contact problem and reveal some interesting features, which were overlooked before.
We show that such a simple example undergoes some surprising transitions in the evolution of contact area, contact perimeter, mean pressure and the
probability distribution of contact pressure.
These results have important implications for analytic models of rough contact,
including asperity based 
models~\cite{bush1975w,greenwood2006w,greenwood1966prcl,mccool1986w,thomas1999b} 
and Persson's contact model~\cite{persson2001jcp,persson2001prl} (for a review, see~\cite{carbone2008jmps,yastrebov2014arxiv}).
We also establish and discuss a closed-form solution for the probability density of the contact pressure for a double sine wave function, which might help revisiting the extension of Persson's model to partial contact~\cite{persson2002prb}.

The paper is organized as follows. In Section~\ref{sec:setupt}, we set up the problem.
In Section~\ref{sec:jgh}, the classical results of Johnson, Greenwood, Higginson are discussed and compared with new numerical
results in Section~\ref{sec:contact_area}. The shape of the contact area is analyzed in Section~\ref{sec:shape}. 
We then discuss the compactness of the contact area shape, the contact perimeter, and reveal associated geometrical 
transitions (Section~\ref{perim}). A detailed study of the contact pressure spatial distribution 
is provided in Section~\ref{sec:pressure}.
In Sections~\ref{sec:pdf} and \ref{sec:persson} the probability distribution of contact pressure is investigated and discussed with respect to Persson's model, respectively.
The results are summarized in Section~\ref{sec:concl}. In Appendix~\ref{app:data} and \ref{app:pdf} additional numerical data 
and the derivation of the probability density for a double sine surface are given.

\section{\label{sec:setupt}Problem set-up and numerical methods}

Let us consider an elastic half-space ($E_1,\nu_1$ are the Young's modulus and Poisson's ratio, respectively) with a wavy surface (Fig.~\ref{fig:1})
\be
  z(x,y) = B \cos\left(2\pi x/\lambda\right)  \cos\left(2\pi y/\lambda\right)
  \label{eq:surface}
\ee
This surface is gradually pressed in frictionless non-adhesive contact with a flat elastic surface ($E_2$, $\nu_2$)
by a pressure $p_0$ applied at infinity. We perform numerical simulations using an FFT-based
boundary element method (BEM)~\cite{stanley1997}\footnote{The original formulation of the FFT algorithm~\cite{stanley1997} contains some errors.
The most relevant is that the solution is shifted in Fourier space by one wavenumber.
For the current and previous studies~\cite{yastrebov2012pre,yastrebov2014arxiv} we use a corrected version of this method, which was validated on many cases.}.
These simulations serve to find the evolution of the contact area, contact perimeter and spatial pressure distribution. 
The simulations are performed on a grid 
of $n\times n = 4096\times4096$ points. The external pressure is applied in 200 load steps up to the pressure $p^*$ ensuring the full contact between surfaces~\cite{johnson1985ijms}
$$
  p^* = \sqrt{2}\pi E^* B/\lambda,
$$
where $E^* = 1/((1-\nu^2_1)/E_1 + (1-\nu^2_2)/E_2)$ is the effective Young's modulus.

We compute the contact area fraction $A'=A/A_0$ or $A'=A/\lambda^2$ as the ratio of points in contact to the total number of points\footnote{The nominal contact area is $A_0=\lambda^2$ as 
we carry out our simulations on a surface $\lambda\times\lambda$ to make use of periodic boundary conditions. In the following, however, due to symmetry all results will be shown only on a quarter of the simulation domain.}.
Computed in this way, the contact area fraction converges to a 
continuous value as $N$ tends to infinity.

\section{\label{sec:jgh}Analysis of Johnson, Greenwood, Higginson}

In~\cite{johnson1985ijms} the authors derive two asymptotic solutions for the considered problem. The first is applicable at infinitesimal contact $p_0/p^* \ll 1$. It uses Hertz theory and neglects elastic interaction between contact spots. The curvature at the crest is 
$4\pi^2B/\lambda^2$, so the relative contact area
at small loads is given by
\be
  A' = \pi\left(\frac{3 p'}{8\pi}\right)^{\!\frac23},
\label{eq:asympt:1}
\ee
where $p'=p_0/p^*$.

Near full contact, when the entire surface is in contact except a small circular region in the deepest valley, the authors suggest an analogy with
a pressurized ``penny-shaped'' crack with no singularity in stress at the edge, it gives
\be
  A' = 1 - \frac{3}{2\pi}\left(1-p'\right)
  \label{eq:asympt:2}
\ee

The same authors carried out numerical simulations using an FFT based boundary element method\footnote{This method is quite similar to the one~\cite{stanley1997} used in this paper,
both are based on the Kalker's variational formulation~\cite{kalker1977jima,kalker1972jem}} and performed real experiments on a silicon rubber block put in contact with a flat surface.
The block dimensions are $80\times80$ mm, the wavelength $\lambda=40$ mm and the amplitude $B=0.24$ mm. Although such an experimental setup does not represent a half-space
with an infinite periodic wavy surface, the obtained experimental results appear to be in good agreement with numerical results and asymptotic estimations~\eqref{eq:asympt:1} and \eqref{eq:asympt:2}. The ensemble of results obtained in~\cite{johnson1985ijms} is depicted in Fig.~\ref{fig:2},a and complemented by recent numerical results~\cite{krithivasan2007tl}.
All points describing the evolution of the contact area seem quite accurate and the authors of~\cite{johnson1985ijms}  judged that the number of data points was sufficient to trace a master curve,
which was assumed to be \emph{everywhere concave}. A similar \emph{concave} master curve can be found in~\cite[Fig. 6]{krithivasan2007tl}.
As shown in Fig.~\ref{fig:2},b, the aforementioned authors underestimated the complexity of the contact area evolution.
Using a larger number of load steps, we reveal that in a certain interval the evolution of the contact area may be a convex function of the contact pressure
and consequently the mean pressure $\bar p=p'/A'$ becomes a non-monotonous function of the external pressure and contact area (see Fig.~\ref{fig:2b}).
Note that a trace of this convexity may be guessed in the coarse data points from~\cite{johnson1985ijms,krithivasan2007tl}, see Fig.~\ref{fig:2},a.

\section{\label{sec:contact_area}Contact area evolution}

\begin{figure}[htb!]
\begin{center}
 \includegraphics[width=0.6\textwidth]{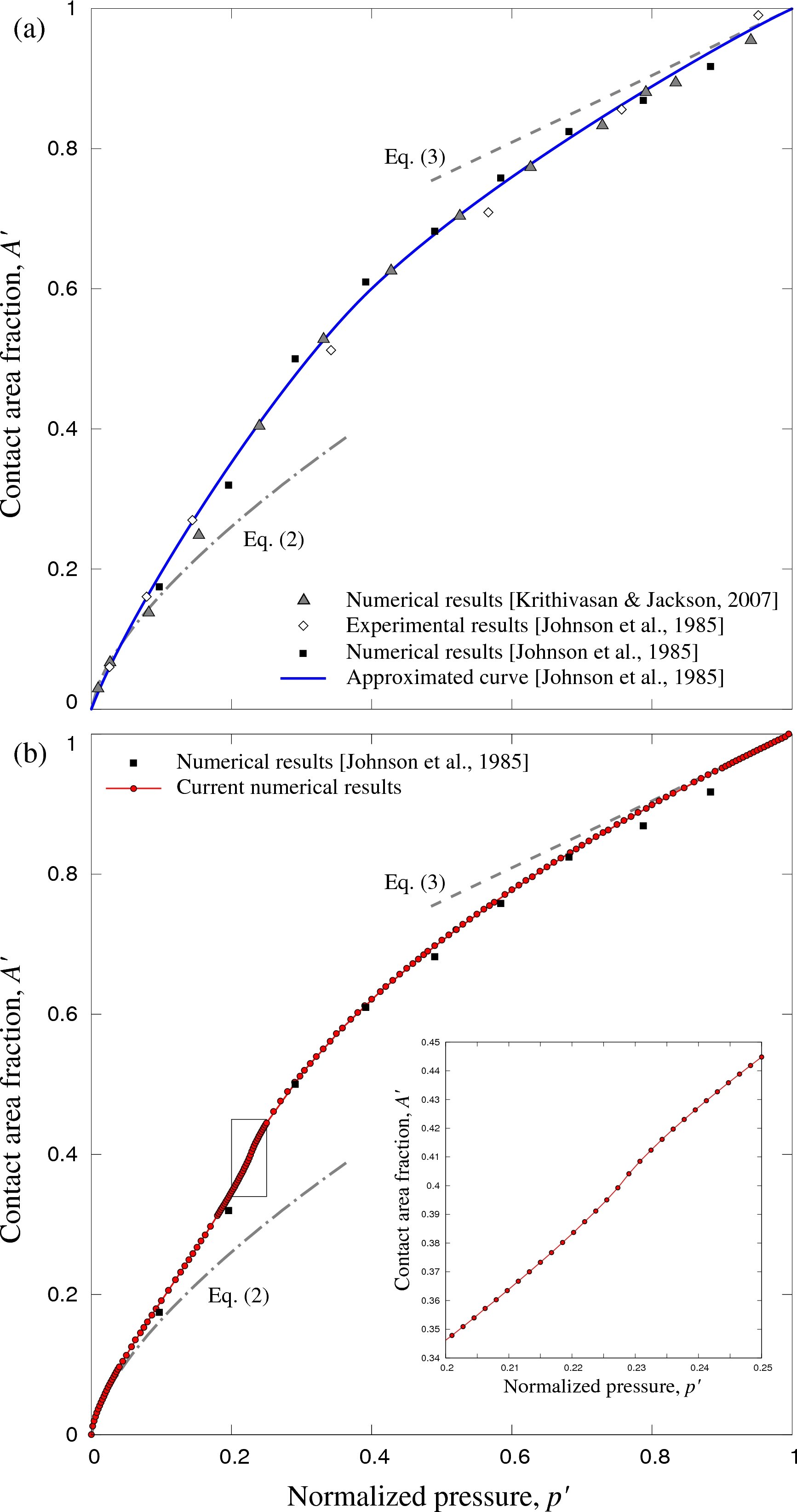}
\end{center}
 \caption{\label{fig:2}Evolution of the contact area: (a) results from the literature~\cite{johnson1985ijms,krithivasan2007tl}; (b) actual numerical results and the zoom on the transition region in the inset}
\end{figure}

\begin{figure}[htb!]
 \includegraphics[width=1\columnwidth]{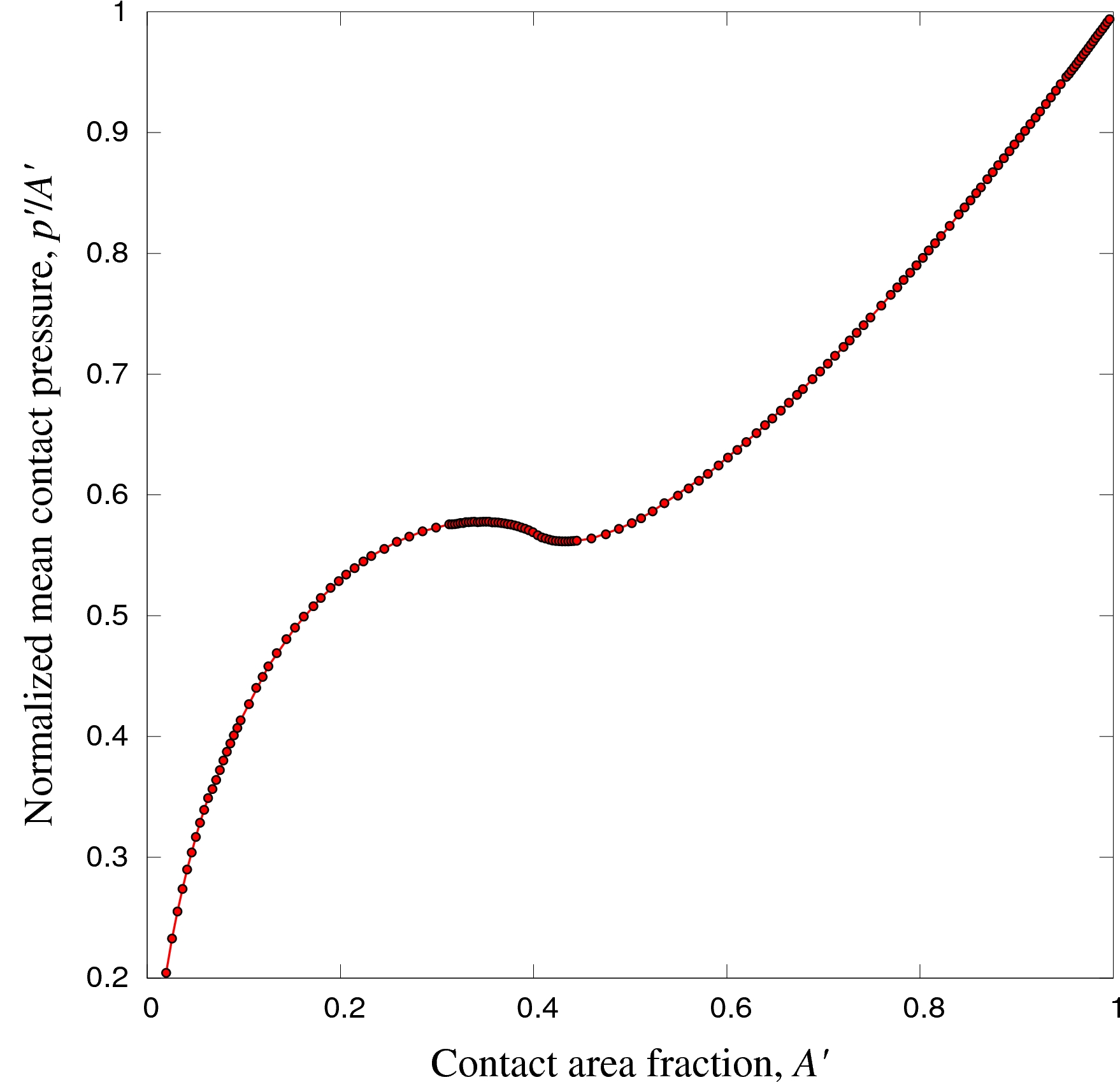}
 \caption{\label{fig:2b}Evolution of the mean contact pressure $p'/A'$ as a function of the contact area fraction}
\end{figure}

\begin{figure}
\begin{center}
 \includegraphics[width=0.6\textwidth]{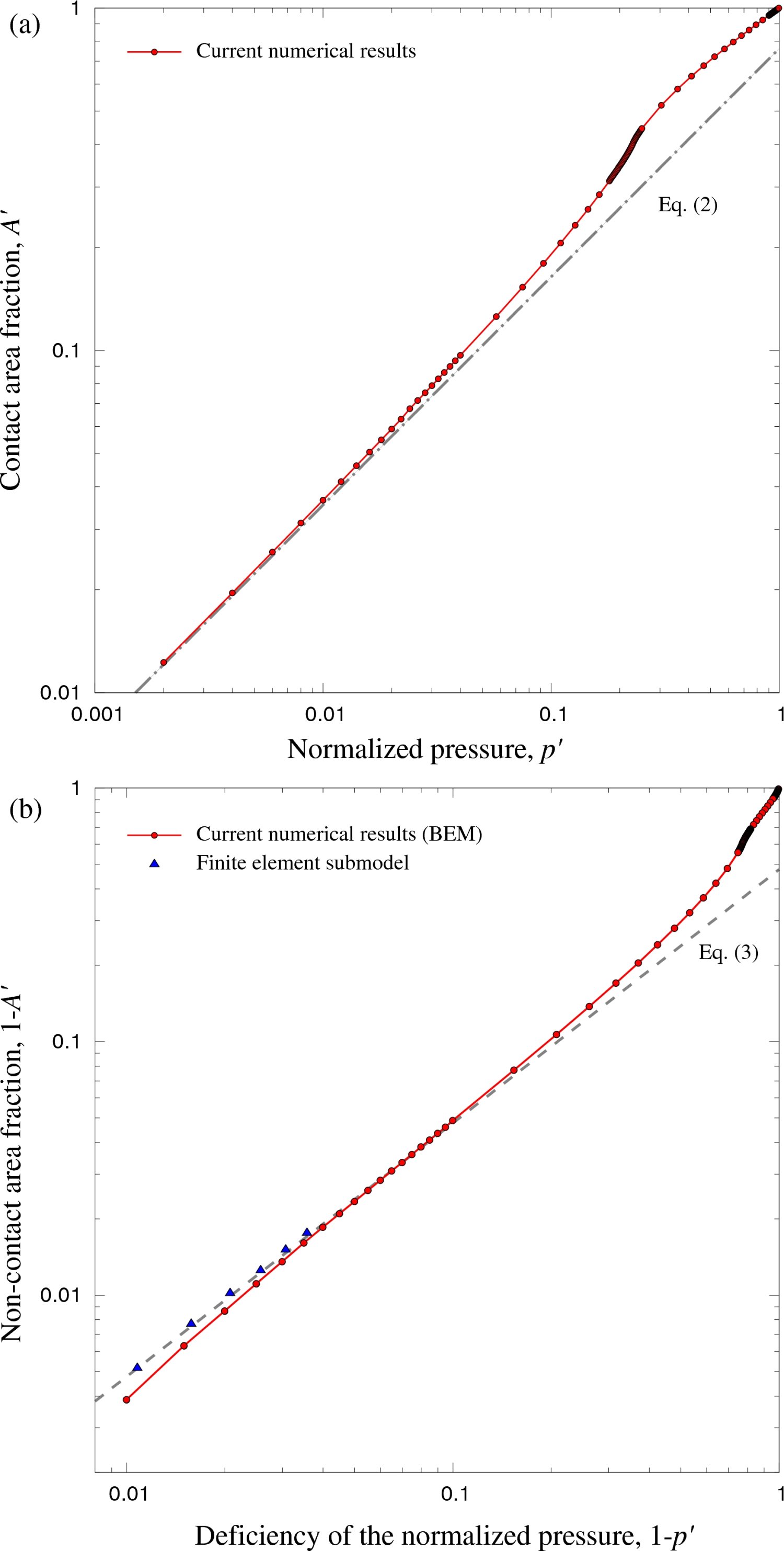}
\end{center}
 \caption{\label{fig:area_zoom}Logarithmic plots of (a) the contact area evolution with pressure compared with the asymptotic solution at small pressures;
and (b) the non-contact area evolution with pressure deficiency compared with pressurized penny-shape crack asymptotic solution near full contact; triangles represent results obtained with an axisymmetric finite element model, which appears to be more accurate near full contact than the FFT-based model}
\end{figure}

The contact area is evidently a monotonous function of the pressure $d A'(p_0)/dp_0 > 0$, which is well approximated by asymptotic \eqref{eq:asympt:1} for infinitesimal and full\footnote{The FFT-based method, which we use, fails to predict accurately the contact area evolution near full contact $A' \gtrapprox 97\%$. Thus, to compare with the asymptotic solution near full contact~\eqref{eq:asympt:2}, we used a more accurate axisymmetric finite element model with discretization $6400$ points per wavelength (triangles in Fig.~\ref{fig:area_zoom},b).} contact by \eqref{eq:asympt:2} (see Fig.~\ref{fig:area_zoom}).

\begin{figure}[htb!]
 \includegraphics[width=1\columnwidth]{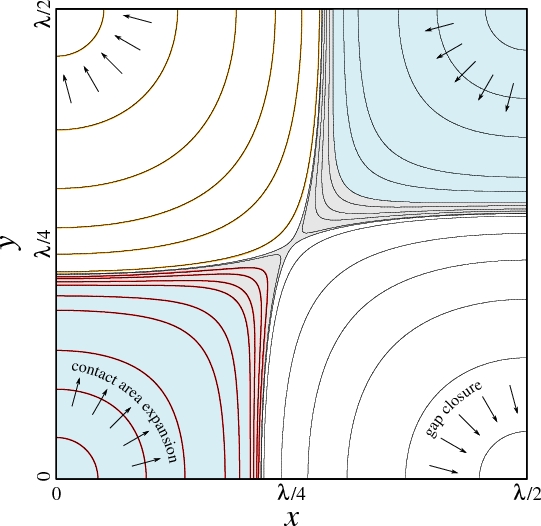}
 \caption{\label{fig:shape}Shape of the contact area at different pressures; the contact area grows from left bottom and right top corners toward the center up to reaching the critical convex shape (shaded in light blue), right after the shape looses convexity up to the reaching the junction (corresponding contact area shaded in gray).
These transitional shapes correspond to the local maximum and minimum in mean contact pressure (see inset in Fig.~\ref{fig:2},b);
black points depict numerical results, solid curves (red and orange) are the fits of Gielis formula~\eqref{eq:superformula2}; fit coefficients determined by
least mean square are given in Table~\ref{tab:coefficients}
}
\end{figure}

As mentioned, in the interval of intermediate contact areas, we find a surprising deviation from the master curve assumed in~\cite{johnson1985ijms} (compare Figs.~\ref{fig:2},a and b).
Inflection points exist, changing the curve from concave to convex and back to concave.
In the considered case the mean contact pressure $\bar p = p_0/A'$ is a monotonously increasing function if and only if the secant of every point of the curve $A'(p')$ is bigger than the tangent at this point, i.e.
\be
  \frac{A'}{p_0} > \frac{dA'}{dp_0}.
  \label{eq:mono_cond}
\ee
This equation is satisfied both for light and high pressure asymptotes, \eqref{eq:asympt:1} and \eqref{eq:asympt:2}, respectively.
It is also evident to see that from Fig.~\ref{fig:2},b. 
The violation of condition \eqref{eq:mono_cond} necessitates inflection points, which can be found by equating the terms in \eqref{eq:mono_cond}.
This change in the evolution of the contact area is connected with two transitions occurring in the growth of the contact zone (see Fig.~\ref{fig:shape}).
The first point corresponds to the moment when the contact area looses convexity and forms a quasi-square shape. 
The second point corresponds to the moment
when two separate contact zones merge. In contrast to the classic geometrical overlap model~\cite{dieterich1996tp}, these transitions occur at different 
pressures (and contact areas). 

We would like to emphasize this surprising behavior of the mean contact pressure.
During the observed transition, the local contact pressure at every contact point increases only slightly, but the contact zone extends rapidly (see Section~\ref{sec:pressure}),
which results in the mean-contact-pressure drop.

\section{\label{sec:shape}Shape of the contact area}

The shape evolution of the contact area (Fig.~\ref{fig:shape}) is strictly asymmetric for area expansion (small pressures) and gap closure (close to full contact).
In the first case, the contact area rapidly looses circular form. 
Its shape can be approximated by the Gielis ``superformula''~\cite{gielis2003generic}, which can be written in the following form in polar coordinates, taking into account all necessary symmetries:
\be
  \varrho(\varphi) = \varrho_0\left[\left|\cos(\varphi)\right|^{k}+\left|\sin(\varphi)\right|^{k}\right]^{-\frac{1}{q}}
  \label{eq:superformula2}
\ee
In Fig.~\ref{fig:shape} we plot the boundary of the contact area at different pressures (see Table~\ref{tab:coefficients} in Appendix A).
Eq.~\ref{eq:superformula2} fits perfectly most of the numerically found boundaries, except right before and after the junction moment. 
Transition states corresponding to convexity change in the contact area evolution are shaded.

\section{\label{perim}Contact perimeter, compactness and percolation limit}

The ratio of the contact perimeter to the square root of the contact area $S'/\sqrt{A'}$ is a measure of compactness of contact spots, where $S'=S/\lambda$ is a normalized perimeter.
Equivalently, near the full contact a more relevant quantity is the measure of compactness of non-contact spots $S'/\sqrt{1-A'}$.
For a general case of $N$ random contact spots with $S'_i, A'_i$ being the relative perimeter and area of $i$-th spot, this ratio is 
$$
  F(N) = \frac{S'}{\sqrt{A'}} = \frac{\sum_i S'_i}{\sqrt{\sum_i A'_i}} = \sqrt{N}\frac{\langle S'\rangle}{\sqrt{\langle A'\rangle}}. 
$$
The ratio $c=F(N)/\sqrt{N}$, that characterizes compactness of contact spot(s) can be easily found for many simple forms:
for a rectangle with ratio of sides $\xi$
$$c_{\mbox{\tiny rectangle}}=2\sqrt{\xi+1/\xi+2},$$
for an ellipse\footnote{One can use Ramanujan's approximation for the perimeter of an ellipse with semi-axes $a$ and $b$, $S\approx\pi(3(a+b)-\sqrt{(3a+b)(a+3b)})$.} with ratio of axes $\xi$
$$c_{\mbox{\tiny ellipse}}\approx\sqrt{\pi}(3\sqrt{\xi+1/\xi+2}-\sqrt{3\xi+3/\xi+10}),$$
for a square and a circle if one puts $\xi=1$:
$$c_{\mbox{\tiny square}}=4,\quad c_{\mbox{\tiny circle}}=2\sqrt{\pi},$$
the circle is the most compact form, so $c=2\sqrt{\pi}$ is the infimum compactness value.

\begin{figure}[htb!]
 \includegraphics[width=1\columnwidth]{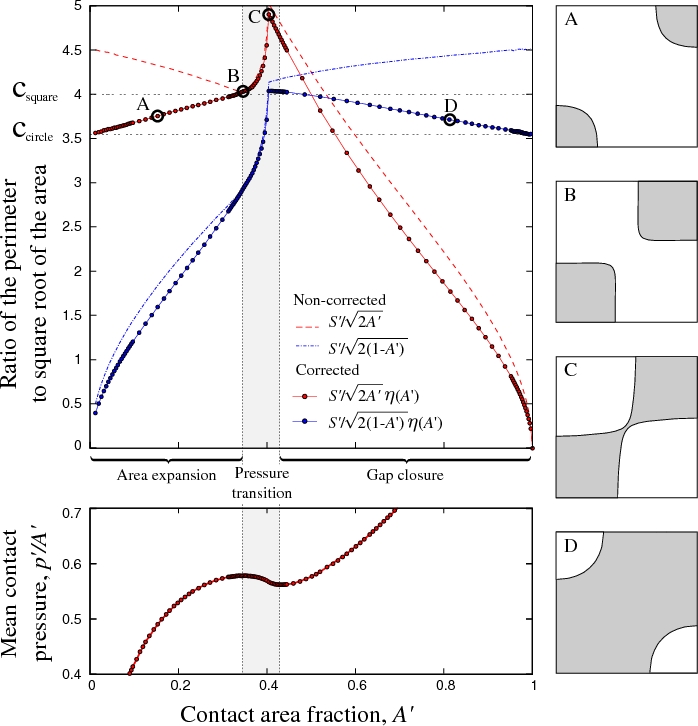}
 \caption{\label{fig:perimeter_area}Evolution of the shape compactness $c=S'/\sqrt{2A'}$ and $c_{\mbox{\tiny closure}}=S'/\sqrt{2(1-A')}$ with increasing contact area;
evolution of $c$ from circular to square-like shape is quite linear as well as the evolution of $c_{\mbox{\tiny closure}}$ from square-like shape to a circular region. The change in compactness is linked with the variation of the mean pressure (see inset in the lower part); the capital letters refer to specific shapes, which are depicted on the right. To demonstrate the importance of the correcting factor $\eta(A')$ (see Eq.~\eqref{eq:correction_factor}), we plot also non-corrected compactness (dashed red and dash-dotted blue lines)}
\end{figure}

The discrete contact perimeter $S^d$ is computed as the number of switches from contact to non-contact and vise versa
along vertical and horizontal lines of computational grid points. Evidently, measured this way, the contact perimeter for any discretization is always
overestimated except for the case of a square contact zone aligned along the direction of discretization.
In case of a circular contact zone, the contact perimeter is overestimated by a factor $4/\pi$, which is
the ratio of the perimeter of a square of side $2a$ to the perimeter of a circle of radius $a$ inscribed in this square\footnote{See also a discussion in~\cite{yastrebov2014arxiv}.}.

The contact zone, being almost circular at light loads, transforms at some stage into an almost square shape aligned along
vertical and horizontal lines of nodes. 
Next the contact areas loose convexity and merge together. Thereafter, the perimeter's shape becomes more and more circular towards the full contact (see Fig.~\ref{fig:shape}).
To take into account the effect of shape alteration on the measurement of the the contact perimeter we introduce
a piece-wise linear correction factor $\eta(A')$ as a function of the contact area fraction
\be
  \eta(A') = \begin{cases}
                \frac{4}{\pi} - \frac{A'}{A'_{sq}}\left(\frac{4}{\pi}-1\right),&\mbox{ if } 0 \le A' < A'_{sq};\\
		1 + \frac{A'-A'_{sq}}{1-A'_{sq}}\left(\frac{4}{\pi}-1\right),&\mbox{ if } A'_{sq} \le A' \le 1,
             \end{cases}
  \label{eq:correction_factor}
\ee
where $A'_{sq}\approx34.8\%$ is the area fraction corresponding to a quasi-square shape of the contact area.
This correction ensures a more accurate estimation of the contact perimeter in its continuum sense.
The corrected normalized perimeter is computed as
\be
  S' =  \frac{S^d}{\eta(A')n},
 \label{eq:corrected_perimeter}
\ee
where $n$ is the number of discretization points per wavelength $\lambda$.
This correction is crucial as can be seen in Fig.~\ref{fig:perimeter_area}, where we plot non-corrected  
$$c^d=\frac{S^d}{n\sqrt{2A'}},$$ 
and corrected compactness
\be
  c=\frac{S^d}{n\eta(A')\sqrt{2A'}}=\frac{S'}{\sqrt{2A'}}.
  \label{eq:compactness1}
\ee
Indeed, the transformation from the initially circular shape into a square-like shape should be reflected by compactness $c$ increasing from $2\sqrt{\pi}$ to $4$.
But for a non-corrected measure an inverse trend is observed, $c^d$ decreases from value $\approx 4.5$ to $4$, which does not reveal the associated change of the shape.
It is evident, however, that for an arbitrary geometry and multiple contact spots of different shapes, a correction function cannot be worked out, thus a discrete 
measurement \color{black}of compactness may be employed~\cite{bribiesca1997cma,bribiesca2008pr}\footnote{The term contact perimeter, which is introduced in these references, should not be 
confused with the contact perimeter employed here. By contact perimeter the author of~\cite{bribiesca1997cma,bribiesca2008pr} understands 
the number of 
boundaries between 
neighboring pixels, 
that form the discrete shape.}.
Note that $\sqrt{2}$ appears in the denominator of Eq.~\eqref{eq:compactness1} 
as for a simulated periodic surface $\lambda\times\lambda$ one entire and four quarters of asperities come in contact; 
at gap closure, we have four half valleys that remain out of contact, so $N=2$.
Note also that for the considered case, to get a relevant measure of compactness one should normalize by the contact area before junction of contact zones (red circles in Fig.~\ref{fig:perimeter_area}) and by the non-contact area after the junction (blue circles in Fig.~\ref{fig:perimeter_area}).

\begin{figure}[htb!]
 \includegraphics[width=1\columnwidth]{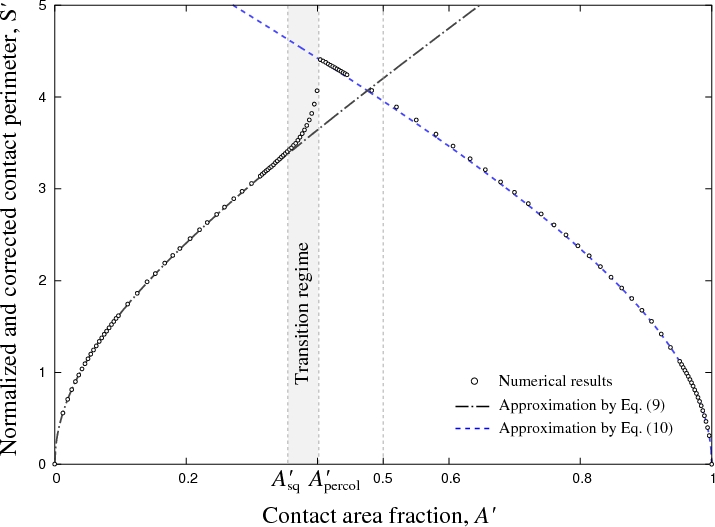}
 \caption{\label{fig:per_evolution}Perimeter evolution and approximations given by Eqs.~\eqref{eq:per_evol_1},\eqref{eq:per_evol_2}; transitional zone from loose of shape convexity to percolation limit is shaded. The central dashed line is drawn to emphasize the asymmetry in the curve}
\end{figure}

The compactness evolves almost linearly from infinitesimal contact state to square-like area shape at $A'_{sq}$, the slope $dc/dA' \approx (4-2\sqrt\pi)/A_{sq}$.
For closing gap, the closure compactness also decreases almost linearly starting from the \emph{percolation area fraction} $A'_{\mbox{\tiny percol}}\approx40.2\%$, 
so\footnote{We found the closure compactness close to the percolation limit to be $c_{\mbox{\tiny closure}}\approx4.04$. Regardless the fact that for ``closing gap'' regime, the shape of the contact boundary cannot be approximated by a square, its compactness measurement near the percolation limit can be with a good accuracy approximated by $c\approx4$.}
$dc/dA' \approx (2\sqrt\pi-4.04)/(1-A'_{\mbox{\tiny percol}})$.
The ratio of these slopes, or simply the ratio of two area transitions, $s\approx A_{sq}/(1-A_{\mbox{\tiny percol}})\approx0.89$, reflects the asymmetry between the initial area evolution and the gap closure.
For a simple geometrical overlap model~\cite{dieterich1996tp}, in which the contact area is found as a cut of a surface geometry by a plane, $s_{\mbox{\tiny geom}}=1$ 
and evidently the percolation limit is $A'_{\mbox{\tiny percol}}=0.5$.
Remark that the \color{black}percolation limit for a simple wavy surface (pressed against an elastic half space) is surprisingly close to the estimation 
$A'_{\mbox{\tiny percol}}\approx42.5\%$ recently obtained in simulations of elastic contact for random fractal surfaces~\cite{dapp2012prl}.

As the compactness $c$ evolves approximately linearly with area, then the contact perimeter may be assumed to evolve as
\be
  S' \!\approx\! 2\sqrt{2A'} \left[\sqrt\pi+\frac{2-\sqrt\pi}{A'_{sq}}A'\right],\;\; A'<A'_{\mbox{\tiny percol}}
  \label{eq:per_evol_1}
\ee
before the percolation limit, and as
\be
  S' \!\approx\! 2\sqrt{2(1\!\!-\!\!A')} \left[\sqrt\pi\!+\!\frac{2.02\!-\!\sqrt\pi}{1\!\!-\!\!A'_{\mbox{\tiny percol}}}(1\!\!-\!\!A')\right],\;\; A'>A'_{\mbox{\tiny percol}}
\label{eq:per_evol_2}
\ee
after this limit. In Fig.~\ref{fig:per_evolution} the evolution of the contact perimeter with contact area is depicted and compared with good accuracy with
Equations~\eqref{eq:per_evol_1} and \eqref{eq:per_evol_2}. 
The switch between two regimes (area expansion and gap closure) is confined within a narrow transition interval confined between the convexity loss and the percolation of contact spots.

The perimeter (compactness) measurements are of interest for characterizing topology of contact zones, adhesion~\cite{pastewka2014pnas}, numerical error estimation~\cite{yastrebov2014arxiv} and for tunneling charge transfer~\cite[e.g., Ch. 23.4]{slade2014b}. 
Moreover, for a random rough surface the rapid increase in compactness may indicate a transition between an asperity based contact state, which implies separate elliptic contact regions, to a more complex state with complex junctions between these regions.

\section{Contact pressure\label{sec:pressure}}

\begin{figure}[htb!]
\begin{center}
  \includegraphics[width=0.8\textwidth]{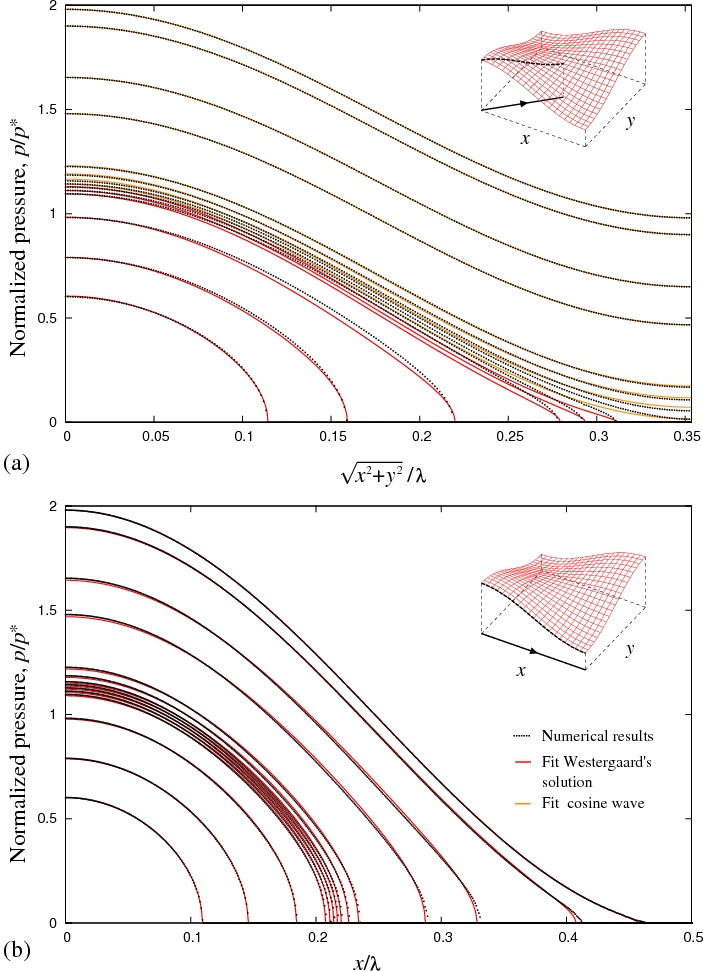}
\end{center}
 \caption{\label{fig:pressure_distribution}Contact pressure distribution along (a) a diagonal line passing through crests ($y=x$) and 
(b) a horizontal line ($y=0$) at different external pressures $p'=3,7,14,20,21,22,23,24,26,29,50,66,90$\%. 
Points correspond to the numerical results; red curves in (a) and (b) are Westergaard's solution, Eq.~\eqref{eq:westergaard}, for corresponding contact radius and maximum contact pressure; orange curves in (a) are simply the fits of a cosine wave. For a better visibility only every fourth point is plotted}
\end{figure}

\begin{figure}[htb!]
\begin{center}
 \includegraphics[width=0.65\columnwidth]{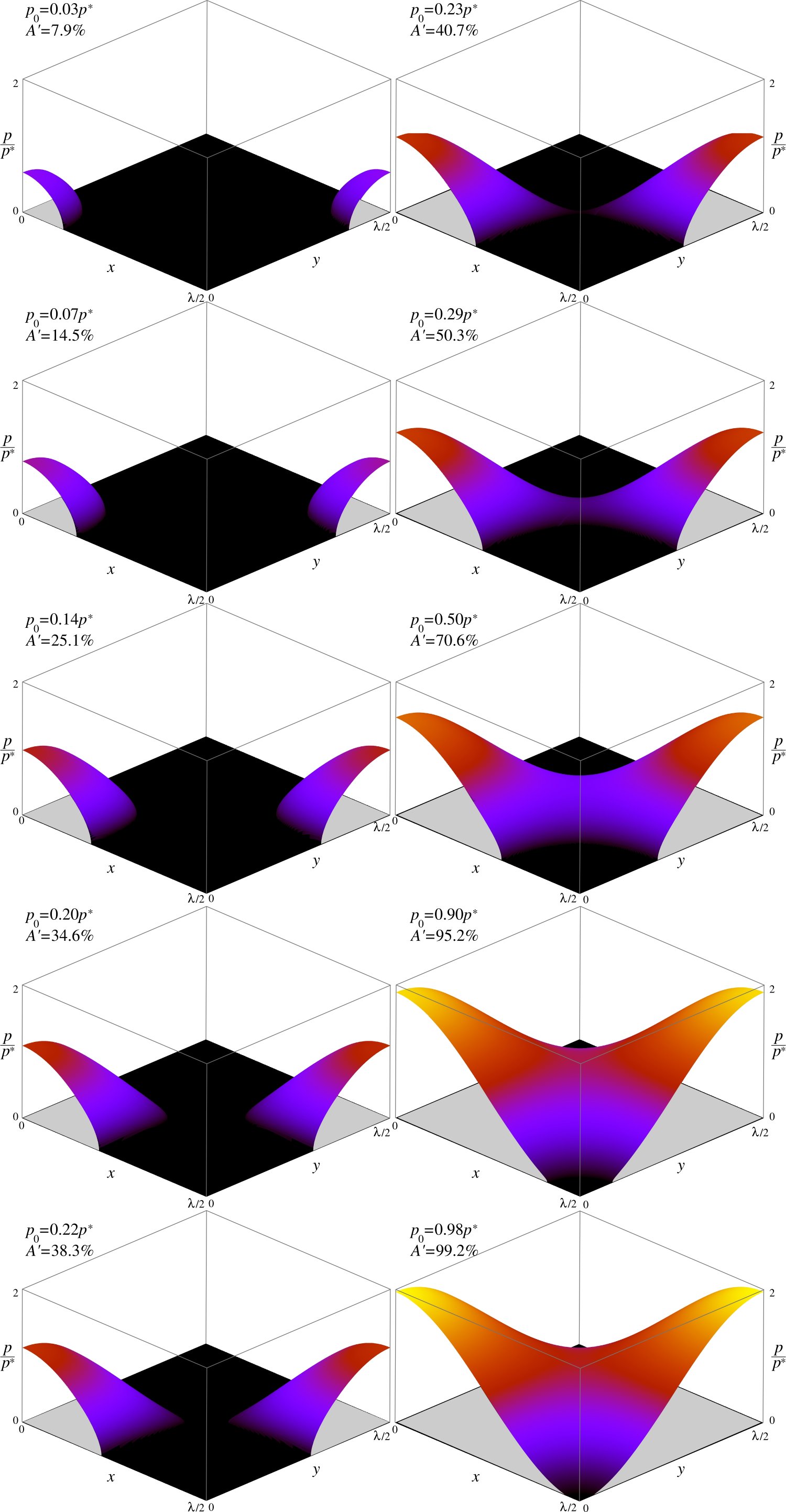}
\end{center}
 \caption{\label{fig:pressure_distribution_3D}Spatial distribution of contact pressure at different loads. The colors aide to visualize the pressure magnitude}
\end{figure}

We plot the contact pressure distribution along a diagonal and a horizontal line in Fig.~\ref{fig:pressure_distribution}, a and b, respectively.
We follow its evolution with increasing pressure. To verify the importance of the two-dimensionality of the problem, we compare
the numerical results along symmetry axes and Westergaard's solution obtained for a one-dimensional sinusoidal profile~\cite{westergaard1939jam}
\be
  p(x,a) = 2 p_0 \frac{\cos\left(\frac{\pi x}{\lambda}\right)}{\sin^2\left(\frac{\pi a}{\lambda}\right)} \sqrt{\sin^2\left(\frac{\pi a}{\lambda}\right)-\sin^2\left(\frac{\pi x}{\lambda}\right)}
  \label{eq:westergaard}
\ee
Surprisingly, this solution fits accurately the contact pressure distribution along the horizontal profile for the entire range of pressures.
It is not quite the case for the pressure distribution along the diagonal line, especially close to the junction between contact zones.
It is worth noting that along this line, the contact pressure rises faster than Westergaard's fit.
The complete spatial distribution of the contact pressure is depicted in Fig.~\ref{fig:pressure_distribution_3D}.

\section{Probability density of contact pressure\label{sec:pdf}}

\begin{figure}[htb!]
 \includegraphics[width=1\columnwidth]{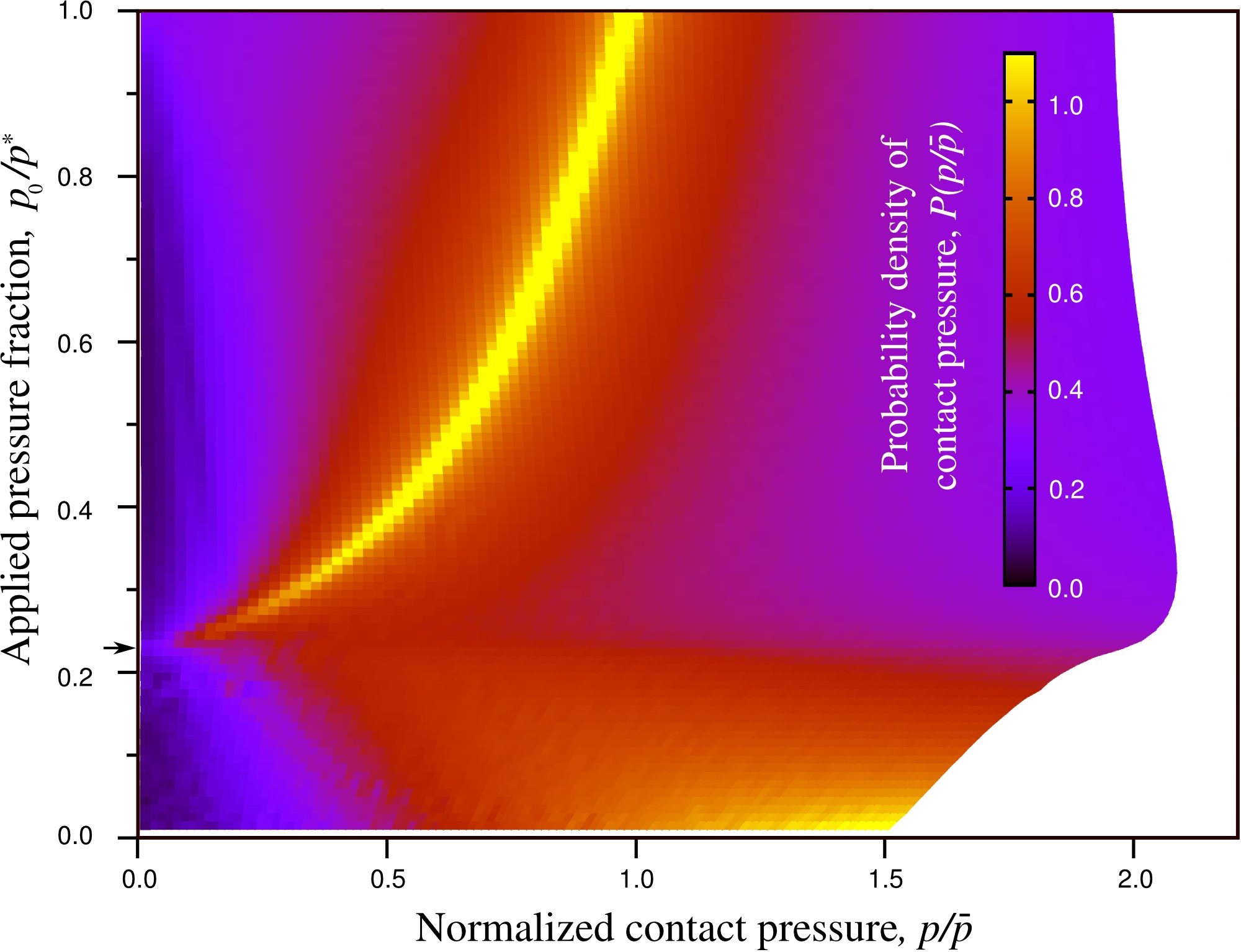}
 \caption{\label{fig:pdf_3D}Evolution of the normalized probability density of normalized contact pressure $P'(\tilde p,p_0)$. Linear evolution at infinitesimal pressures changes to a concave PDF at moderate loads; at reaching the percolation limit $p'\approx0.23$ (marked with an arrow) a singular peak emerges and gradually shifts to higher 
pressure $\tilde p$ at higher loads $p_0$; at full contact the peak position reaches $\tilde p=1$, and there is a non-zero PDF for all pressures including at zero pressure (see Fig.~\ref{fig:pdf_analytics}). }
\end{figure}

In Fig.~\ref{fig:pdf_3D} we plot the evolution of the probability density\footnote{
Hereinafter the PDF of contact pressure is computed only in contact regions; the integral of the PDF over all contact pressures is equal to one.
} (PDF) of normalized contact pressure $P'(\tilde p,p_0)$ under increasing pressure $p_0$, where $\tilde p=p/\bar p$. The PDF experiences a transition separating the regimes before
and after percolation. At the transition a singularity\footnote{We cannot show rigorously that the peak observed in Fig.~\ref{fig:pdf_3D} is a singularity, but we can 
assume that if at full contact a singularity exists, see Eq.~\ref{eq:pdf_full}, it is probable that it also persists at smaller pressures.} 
emerges in the probability density at zero pressure, and moves for increasing pressure $p_0$ towards the center of the distribution (see the bright zone in the figure).

At infinitesimal pressures, when the contact can be approximated by non-interacting Hertzian asperities\footnote{We recall that in this case the contact pressure is $p(r)=p_{\max}\sqrt{1-(r/a)^2}$.}, the PDF of contact pressure of a single asperity can be expressed as
\be
  P(p,p_0 \ll p^*) = \frac{2p}{p_{\max}^2} = \frac{8}{9} \frac{pA'^2}{p_0^2} = \frac{8}{9} \frac{p}{\bar p^2}, 
  \label{eq:pdf_single_asp}
\ee
where the maximal pressure $p_{\max}=3p_0/2A'$. 
The PDF of the normalized contact pressure is 
\be
  P'(\tilde p,p_0 \ll p^*) = \frac89 \tilde p. 
\label{eq:pdf_small}
\ee
At full contact the PDF of contact pressure is simply proportional to the PDF of a wavy surface. However, we could not find in the literature a formula for this quantity. 
We derived that at full contact (see appendix~\ref{app:pdf}) 
\be
  P'(\tilde p,p_0=p^*) = 
\frac{2}{\pi^2}
\frac{\mathrm{F}\left[\arccos(|1-\tilde p|),
\frac{1}{\sqrt{2\tilde p - \tilde p^2}}
\right]}{\sqrt{2\tilde p - \tilde p^2}},
\label{eq:pdf_full}
\ee
where 
$$
\mathrm{F}(l,k)=\int\limits_0^l \frac{1}{\sqrt{1-k^2\sin^2(\phi)}} d\phi
$$
is the incomplete elliptic integral of the first kind.

To evaluate Eq.~\eqref{eq:pdf_full} at $\tilde p = 2$ or equivalently at $\tilde p=0$, 
we replace the variable $\tilde p = 1+\cos(\varphi)$, so $\tilde p \to 2-$ when $\varphi\to0_{+}$, thus Eq.~\eqref{eq:pdf_full} can be rewritten as
$$
P'(\tilde p,p_0\!=\!p^*) \!=\! 
\frac{2}{\pi^2}\frac{\mathrm{F}\left(\varphi,\!1/\sin(\varphi)\right)}{\sin(\varphi)} \xrightarrow[\varphi\to0_{+}]{}\! \frac{2}{\pi^2}\frac{\mathrm{F}\left(\varphi,\!1/\varphi\right)}{\varphi},
$$
where the elliptic integral for $\varphi\to0_{+}$ can be approximated as:
$$
 \mathrm{F}\left(\varphi,1/\varphi\right)\approx\int\limits_0^\varphi \frac{1}{\sqrt{1-t^2/\varphi^2}}\,dt = 
\varphi\int\limits_0^1 \frac{1}{\sqrt{1-s^2}}\,ds = \varphi\frac\pi2.
$$
Thus the PDF of the minimal and maximal contact pressures in full contact is 
\be
  P'(0,p^*) = P'(2,p^*) = 1/\pi.
  \label{eq:as_full_pdf}
\ee
Note that a general result may be obtained by computing the PDF only in vicinity of the peak pressure, where it has a curvature $4\pi^2 p^*/\lambda^2$ yielding
$P=N\lambda^2/(A_0 2\pi p^*)$, where $N$ is the number of asperities per normalization area $A_0$. In considering case the normalized amplitude $p^*/\bar p=1$, $A_0=\lambda^2$ and $N=2$, which also gives $P(0)=1/\pi$. 

\begin{figure}[htb!]
 \includegraphics[width=1\columnwidth]{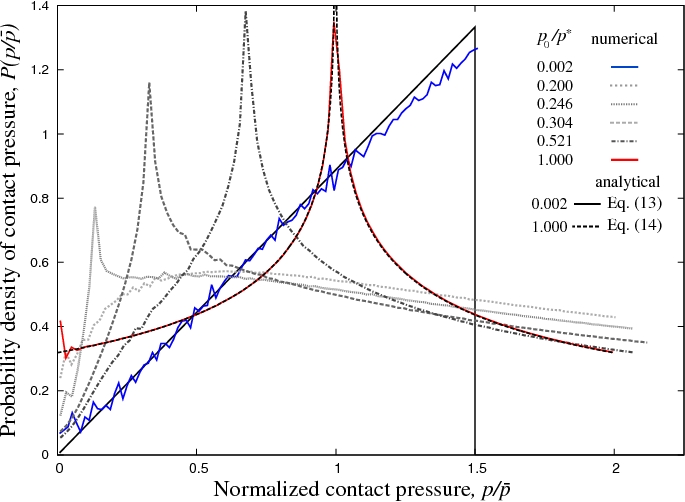}
 \caption{\label{fig:pdf_analytics}Normalized probability density of normalized contact pressure $P'(\tilde p)$ for different external pressures $p_0$. Numerical results at light contact $p_0=0.002p^*$ and $A'\approx1.2\%$ are in good agreement with analytical prediction~\eqref{eq:pdf_small}; at full contact the numerical results match Eq.~\ref{eq:pdf_full}. The singularity peak appears in the PDF at the percolation limit, see also bright color region in Fig.~\ref{fig:pdf_3D}}
\end{figure}

In Fig.~\ref{fig:pdf_analytics} we plot the PDF of contact pressure at different loads. 
At light pressures, $p_0\ll p^*$, the PDF has to be a linear function of the contact pressure. For $p_0=0.002p^*$, our numerical results are in good agreement with analytical prediction~\eqref{eq:pdf_small}.
A small deviation is observed for higher values of local pressure $\tilde p$. The numerical results are slightly noisy due to a relatively small number of contact points ($\approx 206\,000$). 
At higher pressures $p_0$ the PDF looses monotonicity, and, as commented earlier, at reaching the percolation limit $A_{\mbox{\tiny percol}}$, a singular peak emerges at zero pressure $\tilde p=0$, which gradually shifts with increasing pressures towards $\tilde p=1$, which is reached at full contact. 
We showed, Eq.~\eqref{eq:as_full_pdf}, that at full contact the PDF of zero pressure is non-zero $P'(0)=1/\pi$ and it is in perfect agreement with numerical results.
We suppose that 
the PDF may be also non-zero 
at percolation, when contact zones just start to merge and the PDF singularity emerges at zero pressure,
however it is not trivial to prove it analytically nor numerically.

\section{\label{sec:persson}Implications for Persson's model of rough contact}

In contrast to asperity based models \cite{bush1975w,carbone2008jmps,greenwood2006w,greenwood1966prcl}, Persson's model of rough contact 
does not rely on the notion of asperities~\cite{persson2001prl,persson2001jcp,yang2008jpcm}.
The author starts from a full contact between surfaces under external pressure $p_0$; if the surfaces are perfectly flat, the PDF of contact pressure is simply a Dirac-delta function
$P(p)=\delta(p-p_0)$. When the roughness with wavenumbers from $k_1$ to $\zeta k_1$ is gradually introduced in the surface spectrum by increasing $\zeta$, 
then the PDF of contact pressure spreads out and takes a 
Gaussian shape. Note that the full contact is preserved for any $\zeta$.
To describe this evolution, Persson deduced a diffusion equation for the PDF (acts as density of diffusing medium) of contact pressure (acts as spatial coordinate) depending
on the variance of the surface roughness (acts as time):
\be
\ddp{P(p,\zeta)}{V(\zeta)} = \frac{1}{2}\ddps{P(p,\zeta)}{p},
\label{eq:persson_pdf}
\ee
where 
$$V(\zeta) = \frac12E^*m_2(\zeta) = \frac{\pi E^*}{2}\int\limits_{k_1}^{\zeta k_1} k^3\Phi^p(k) dk$$
is the variance of the contact pressure depending on the magnification parameter $\zeta$, which controls the breadth of the surface spectrum (see, e.g.\cite{nayak1971tasme,yastrebov2014arxiv}), $m_2$ is the second spectral moment, $E^*$ is the effective Young's modulus~\cite{johnson1987b}, 
$\Phi^p(k)$ is the radial power spectral density and $k$ is the wavenumber (see~\cite{manners2006w} for a comprehensive derivation of this equation).
However, as the considered surface is rigorously Gaussian, an infinite pressure is needed to maintain full contact for any $\zeta$.
To extend this model to finite pressures and partial contacts it was suggested to impose a boundary condition 
to Eq.~\eqref{eq:persson_pdf}, which postulates that the PDF of zero pressure is always zero~\cite{persson2002prb}
\be
  P(p=0,\zeta) = 0.
  \label{eq:persson_bc}
\ee
This condition seems reasonable in the context of asperity based models: non-interacting asperities, which contact only at their tips of constant curvature (circular or elliptic). In this case, the slope of the contact pressure tends to infinity at contact edges, which ensures the PDF at each asperity of the form~\eqref{eq:pdf_single_asp} 
and validates the boundary condition~\eqref{eq:persson_bc}~\cite{hyun2007ti,manners2006w}. 
In Persson's model, however, one moves from the full contact towards partial contact by decreasing the external pressure to finite values.
In this process the local contact pressure reaches zero at 
valleys 
before they loose the contact. But when a valley is ready to escape contact, locally the PDF of contact pressure
in the limit of zero pressure is similar to a PDF of a wavy surface at full contact Eq.~\eqref{eq:pdf_full}. Thus the PDF of contact pressure at every contact opening must be
non-zero, and the boundary condition~\eqref{eq:persson_bc} may not be fully justified. 
Moreover, we suggest that this perturbation in the boundary condition exists not only at opening valleys but also at any junction between contact spots. 
Since in Persson's model the roughness spectrum
is continuous, for any pressure, an infinite number
of opening points exists, whose density depends on Nayak's parameter~\cite{nayak1971tasme}, root mean squared surface gradient, effective elastic modulus and applied pressure. 
We suggest that Persson's model could be strengthened by replacing the boundary
condition~\eqref{eq:persson_bc} by a pressure dependent positive function.

\section{Conclusion\label{sec:concl}}

 Revisiting a problem of squeezing an elastic wavy surface, which seemed to have been thoroughly addressed about three decades ago~\cite{johnson1985ijms},
 we discovered several notable transitions in mechanical and geometrical quantities.
 These transitions are connected with the shape change of the contact area.
 In particular, the loss of shape convexity and the consequent merge of contact zones result in the local maximum and minimum of the mean contact pressure, respectively.

 The percolation limit, at contact area fraction $\approx 40.2\%$, separates two different regimes in the shape of the probability density (PDF) of contact pressure: without and with a singular peak.
 We found that at full contact the PDF is described by an incomplete elliptic integral of the first kind, and that
 the value of the PDF of zero pressure is non-zero. We suggested that a finite probability of zero pressure may also exist at junctions between contact zones.
 This has important implications for contact of random rough surfaces, for which any detachment point (and possibly any junction point) results in a non-zero PDF of zero pressure.
This finding might help readdress the boundary condition used in
the extension of Persson's model, which assumes a zero PDF at zero pressure for partial contact.
%

We analyzed as well the perimeter and compactness of the shape of contact area, which displayed interesting transitions close to the percolation limit.
These are delicate to measure due to the pixelated shapes obtained in numerical simulations and experimental measurements.
Thus specific techniques and correction factors have to be developed for this purpose.

It will be of interest to reiterate the analysis for different geometries and to consider non-linear materials, adhesive and frictional forces.

\section{Acknowledgment}
We are grateful to James A. Greenwood and to anonymous reviewers for constructive comments.
GA and JFM greatly acknowledge the financial support from the European Research Council (ERCstg UFO-240332).

\appendix

\section{\label{app:pdf}Probability density function of a wavy surface}

\begin{figure}
\centering \includegraphics[width=0.8\columnwidth]{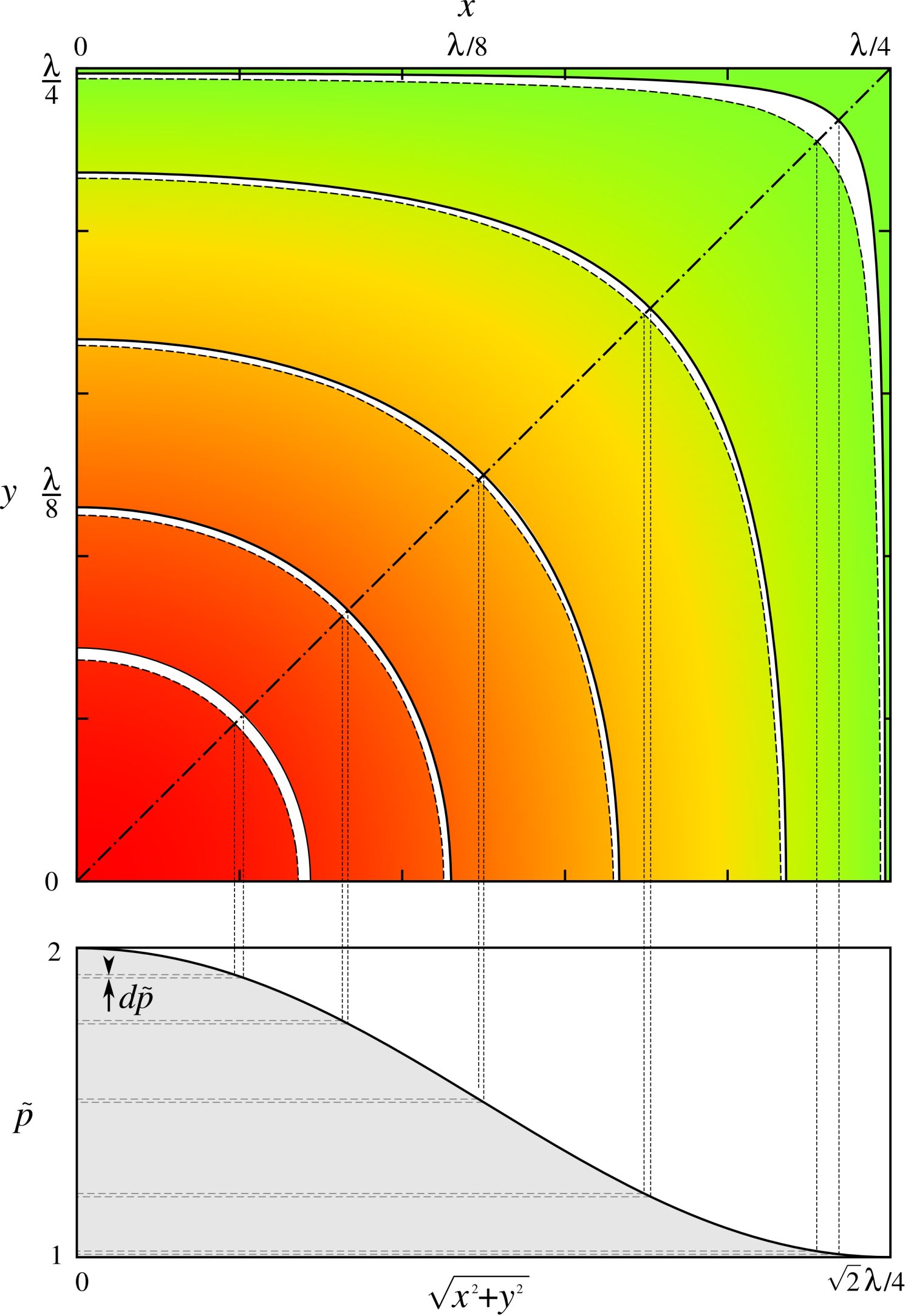}
 \caption{\label{fig:pdf_definition}Schematic figure explaining the computation method for the probability density function of a wavy surface. Iso-pressure curves $y(x,\tilde p)$ and $y(x,\tilde p+dp)$ at different pressures $\tilde p$, the area between curves is proportional to the probability density increment $P(\tilde p,\tilde p+dp)dp$}
\end{figure}

\begin{figure}
 \includegraphics[width=1\columnwidth]{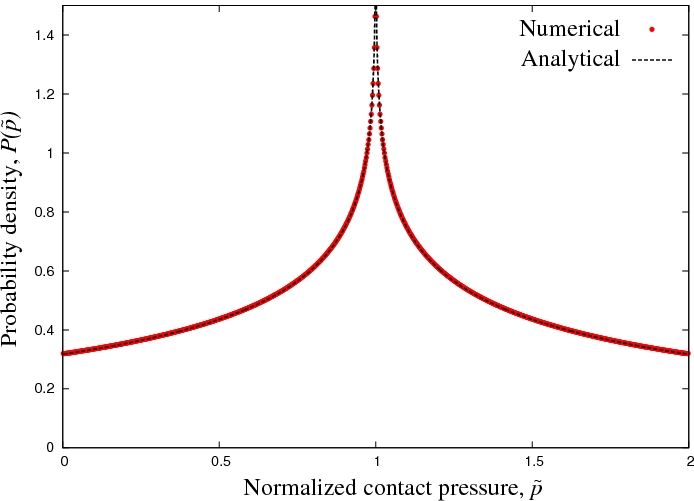}
 \caption{\label{fig:pdf_comparison}Comparison of the analytically evaluated~\eqref{eq:pdf_3D_} and numerically evaluated ($2\cdot10^8$ grid points per $\lambda/4\times\lambda/4$ and 500 bins) probability density function of a wavy surface}
\end{figure}

To obtain the PDF of the contact pressure for the case of full contact 
$$
  \tilde p = 1+\cos\left(2\pi x/\lambda\right)  \cos\left(2\pi y/\lambda\right)
$$
we use the following observation.
If for a given pressure we express the iso-pressure curve as $y=y(x,\tilde p)$, then the area between two iso-curves $y(x,\tilde p)$ and $y(x,\tilde p+dp)$ would
be proportional to the increment of the probability density of the contact pressure $P(\tilde p,\tilde p+dp)dp$ (see Fig.~\ref{fig:pdf_definition}). At the full contact 
$\tilde p \in [0,2]$. Then the iso-pressure curve is given by: 
\be
  y=\frac{\lambda}{2\pi}\arccos\left(\frac{\tilde p-1}{\cos(2\pi x/\lambda)}\right).
\ee
The area under this curve in the range $x,y\in[0;\lambda/4]$, i.e. for $\tilde p\in[1;2]$ may be expressed as:
\be
  A(\tilde p) = \int\limits_{0}^{\frac{\lambda}{2\pi}\arccos(\tilde p-1)} \frac{\lambda}{2\pi}\arccos\left(\frac{\tilde p-1}{\cos(2\pi x/\lambda)}\right)\,dx
  \label{eq:area_isopressure}
\ee
The differential of the probability density is then 
$$P(\tilde p,\tilde p+d p) dp = \frac{A(\tilde p)-A(\tilde p+d p)}{A_0},$$
where $A_0=\lambda^2/16$ is the considered area. In the limit $dp\to0$ we obtain
$$P(\tilde p) = -\frac{1}{A_0}\frac{dA(\tilde p)}{d\tilde p}$$
Substituting the integral form~\eqref{eq:area_isopressure} in this expression enables us to evaluate this derivative using the differentiation under the integral sign\footnote{Differentiation under the integral sign: 
$$\frac{d}{dx}\!\!\!\int\limits_{a(x)}^{b(x)}\!\!\!\!f(x,t)dt =
\!f(x,b(x))\frac{db(x)}{dx} - f(x,a(x))\frac{da(x)}{dx} + \!\!\!\int\limits_{a(x)}^{b(x)} \!\!\frac{df(x,t)}{dx}dt
$$
}
\be
 \begin{split}
 &\frac{dA(\tilde p)}{d\tilde p}= \frac{\lambda}{2\pi}\int\limits_{0}^{\frac{\lambda}{2\pi}\arccos(\tilde p-1)} \frac{dx}{\cos(2\pi x/\lambda)\sqrt{1-\frac{(\tilde p-1)^2}{\cos^2(2\pi x/\lambda)}}} =\\
&=-\frac{\lambda}{2\pi\sqrt{2\tilde p -\tilde p^2}}\int\limits_{0}^{\frac{\lambda}{2\pi}\arccos(\tilde p-1)} \frac{dx}{\sqrt{1-\frac{\sin^2(2\pi x/\lambda)}{2\tilde p - \tilde p^2}}} =\\
&=
-\frac{\lambda^2}{4\pi^2\sqrt{2\tilde p -\tilde p^2}}\int\limits_{0}^{\arccos(\tilde p-1)} \frac{d\varphi}{\sqrt{1-\frac{\sin^2(\varphi)}{2\tilde p - \tilde p^2}}}=\\
&=-\frac{\lambda^2}{4\pi^2\sqrt{2\tilde p -\tilde p^2}}\;\mathrm{F}\left(\arccos(\tilde p-1),\frac{1}{\sqrt{2\tilde p - \tilde p^2}}\right),
 \end{split}
\ee
where $\mathrm F(l,k)=\int\limits_0^l 1/\sqrt{1-k^2\sin^2(\varphi)}\,d\varphi$ is the incomplete elliptic integral of the first kind. 
So the probability density of the contact pressure for $\tilde p \in [1;2]$ or equivalently for $\{x,y\}\in[0;\pi/4]$ is
$$
  P(\tilde p) = \frac{4}{\pi^2}\frac{\mathrm{F}\left(\arccos(\tilde p-1),1/\sqrt{2\tilde p - \tilde p^2}\right)}{\sqrt{2\tilde p - \tilde p^2}}
$$
To extend it to a periodic domain $x,y\in \mathbb R$ and for $\tilde p\in[0;2]$, one needs simply to take the absolute value of the argument in $\arccos$ and divide the PDF by a factor of two
\be
  P(\tilde p) = \frac{2}{\pi^2}\frac{\mathrm{F}\left(\arccos(|1-\tilde p|),1/\sqrt{2\tilde p - \tilde p^2}\right)}{\sqrt{2\tilde p - \tilde p^2}}
  \label{eq:pdf_3D_}
\ee
This expression is depicted in Fig.~\ref{fig:pdf_comparison} and compared with numerically evaluated probability density of a wavy surface computed on the grid of $2\cdot10^8$ equally spaced points in the region $\{x,y\}\in[0;\pi/4]$ using 500 bins.

\section{Data\label{app:data}}

In Table~\ref{tab:coefficients} some numerical results are presented (pressure, area and perimeter)
as well as coefficients of Gielis formula~\eqref{eq:superformula2} that fit the corresponding area shape.

\begin{table}[htb!]
\begin{center}
\begin{tabular}{ccccccc}
  $p' (\%)$ &  $A' (\%)$ &  $p'/A'$ & $S'$ & $q$ & $k$ & $\rho_0$\\\hline
0.20&	1.23&	0.163&	0.558&	128.439& 3.857&	0.177\\
2.00&	5.90&	0.339&	1.244&	12.374&	3.081&	0.382\\
5.75&	12.55&	0.458&	1.861&	7.520&	3.524&	0.547\\
12.75&	23.20&	0.549&	2.632&	6.285&	4.750&	0.718\\
16.25&	28.51&	0.570&	2.972&	6.929&	6.094&	0.779\\
19.23&	33.29&	0.577&	3.261&	9.001&	8.927&	0.825\\
  \multicolumn{7}{l}{* Contact area looses convexity $p'\approx20.1\%$, $A_{\mbox{\tiny sq}}\approx34.8\%$}\\
20.28&	35.09&	0.578&	3.393&	10.946&	11.360&	0.839\\
21.33&	37.00&	0.576&	3.529&	15.601&	17.117&	0.852\\
22.03&	38.37&	0.574&	3.690&	-&	-&	-\\
22.73&	39.93&	0.569&	4.068&	-&	-&	-\\
  \multicolumn{7}{l}{* Percolation limit $p'\approx22.8\%$, $A_{\mbox{\tiny percol}}\approx40.2\%$}\\
22.90&	40.41&	0.567&	4.407&	12.511&	9.673&	0.707\\
23.78&	42.31&	0.562&	4.331&	8.666&	6.668&	0.699\\
30.42&	52.01&	0.585&	3.891&	6.607&	4.365&	0.652\\
41.25&	63.23&	0.652&	3.327&	7.112&	3.671&	0.582\\
52.08&	75.99&	0.685&	2.606&	9.397&	3.267&	0.478\\
79.17&	89.38&	0.886&	1.677&	19.167&	3.034&	0.322\\
96.50&	98.40&	0.981&	0.637&	127.513& 2.895&	0.126\\
\hline
\end{tabular}
\end{center}
\caption{\label{tab:coefficients}Evolution of the contact area fraction $A'$ and corrected perimeter $S'$ (see Eq.~\eqref{eq:corrected_perimeter}) with increasing pressure $p'=p_0/p^*$;  $q,k,\rho_0$ are the coefficients of Gielis formula~\eqref{eq:superformula2} that approximate the shape of the contact perimeter, coefficients are found by the least mean error fit.}
\end{table}


\end{document}